\documentclass[aps,prl,twocolumn,aps,floatfix,showpacs,showkeys, longbibliography, superscriptaddress]{revtex4-2}
\usepackage{amssymb}
\usepackage{amsbsy}
\usepackage{amsmath}
\usepackage{epsfig}
\usepackage{color}
\usepackage{float}
\usepackage{xr-hyper}
\usepackage{bm}
\usepackage[colorlinks,linktocpage,bookmarks=false,citecolor=blue,linkcolor=red,urlcolor=blue]{hyperref}

\newcommand{\beg}{\begin{gather}}
\newcommand{\eeg}{\end{gather}}
\newcommand{\beq}{\begin{equation}}
\newcommand{\eeq}{\end{equation}}
\newcommand{\bea}{\begin{eqnarray}}
\newcommand{\eea}{\end{eqnarray}}

\newcommand{\ket}[1]{\left| #1 \right\rangle}

\newcommand{\be}{\begin{equation}}
\newcommand{\ee}{\end{equation}}
\def\ba{\begin{aligned}}
\def\ea{\end{aligned}}
\def\bes{\begin{subequations}}
\def\ees{\end{subequations}}
\def\bal{\begin{align}}
\def\eal{\end{align}}

\usepackage{soul}
\usepackage{ulem} 
\usepackage{hyperref}
\begin{document}

\title{Spin Chain Quantum Communication on a Trapped-Ion Processor}
\author{Madhumita Sarkar}
\thanks{sarkar.madhumita770@gmail.com}
\author{Trinity Pointon}
\author{Sougato Bose}
\affiliation{Department of Physics and Astronomy, University College London, Gower Street, WC1E6BT, London}
\date{\today}

\begin{abstract}
Efficient communication between distant qubits is one of the central challenges in scaling quantum processors. Although engineered spin chain protocols have been extensively investigated theoretically, their experimental realization has remained comparatively limited. Here, we experimentally realize engineered quantum communication protocols through digitally simulated spin Hamiltonian on IonQ's Forte 1/ Forte Enterprise 1 trapped-ion quantum processor. Combining exact numerical simulations with quantum hardware experiments, we benchmark uniform nearest-neighbour and engineered coupling profiles and demonstrate that engineered interactions significantly enhance the fidelity of quantum state transfer. We further show that exploiting the commutation structure of the spin Hamiltonian enables a parallel Trotter decomposition that more faithfully reproduces the target dynamics while substantially reducing the circuit depth and execution time compared to the conventional sequential implementations. Our results demonstrate that programmable quantum processors can effectively realize and efficiently implement quantum communication protocols, bringing Hamiltonian-based quantum communication closer to practical quantum technologies.
\end{abstract}
\maketitle

\section{Introduction}

Efficient communication between distant qubits is emerging as one of the principal challenges in scaling programmable quantum processors. As quantum processors continue to grow in size and complexity, transporting quantum information across a processor with minimal communication overhead becomes increasingly important for scalable quantum computation~\cite{DiVincenzo2000,NielsenChuang2010,Monroe2014}. Unlike classical information, arbitrary quantum states cannot simply be copied because of the no-cloning theorem~\cite{Wootters1982}. Consequently, quantum communication requires protocols that faithfully preserve the encoded quantum information throughout the transfer process.

On gate-based quantum processors, communication between distant qubits is conventionally achieved through sequences of nearest-neighbour SWAP gates. Although universally applicable, this strategy becomes increasingly inefficient as the communication distance grows, since each additional SWAP increases the circuit depth, execution time, and accumulated gate errors. These communication overheads are expected to become an important limitation for large-scale quantum processors, motivating the development of alternative protocols~\cite{Cirac1997,Monroe2014}.

One promising approach is to exploit the intrinsic dynamics of interacting many-body systems. In 2003 it was first proposed that spin chains could function as quantum data buses, allowing quantum information to propagate through the natural Hamiltonian evolution of coupled spins without requiring local control over every intermediate qubit~\cite{Bose2003}. This proposal stimulated extensive theoretical research demonstrating that engineering the coupling profile of a spin chain provides a powerful means of controlling quantum transport, enabling high-fidelity quantum state transfer and a broad range of quantum communication protocols~\cite{Christandl2004,Bose2007,Bose2014,Albanese2004,Yung2005,Shi2005,DiFranco2008,Yao2011,Kay2010}. These developments established engineered spin chains as a promising framework for realizing efficient quantum communication through controlled many-body dynamics.

\begin{figure}[h]
    \centering
    \includegraphics[width=0.98\linewidth]{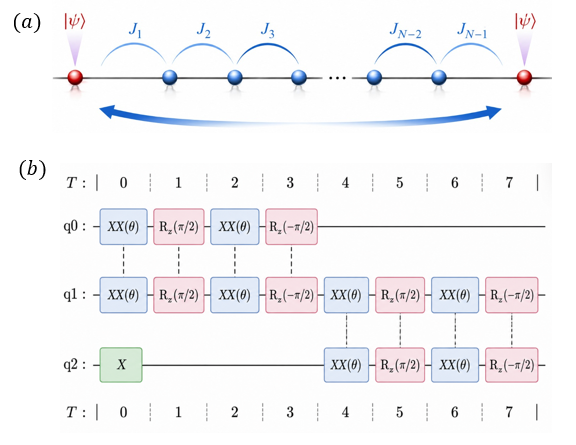}
    \caption{(a) Schematic illustration of quantum state transfer in an engineered spin chain. A quantum state initially encoded in the leftmost qubit, propagates coherently through the chain under the engineered nearest-neighbour couplings {$J_i$} and is reconstructed at the opposite end after the optimal transfer time. (b) Digital implementation of the corresponding nearest-neighbour XY Hamiltonian on the trapped-ion quantum processor. The continuous Hamiltonian evolution is approximated using a first-order Trotter decomposition, in which each Trotter step is realized through a sequence of experimentally accessible two-qubit interactions and single qubit rotations.}
    \label{fig1}
\end{figure}

Despite these significant theoretical advances, experimental realizations of engineered spin chain communication remain comparatively scarce, with demonstrations largely limited to nuclear magnetic resonance systems, photonic waveguide arrays and superconducting circuits~\cite{Zhang2005,PerezLeija2013,PhysRevApplied.10.054009,Jurcevic2014}. This motivates the exploration of programmable quantum computing platforms capable of implementing and benchmarking engineered communication protocols under realistic experimental conditions. Among these, trapped-ion quantum processors provide a particularly attractive architecture owing to their high-fidelity gate operations, all-to-all qubit connectivity, and the flexibility to digitally synthesize programmable spin Hamiltonians~\cite{Kielpinski2002,Blatt2012,Wright2019,Pino2021,Monroe2021}. 

Here, we experimentally investigate engineered quantum communication protocols through the digital simulation of spin Hamiltonians on IonQ's Forte 1/ Forte Enterprise 1 (depending on availability) trapped-ion quantum processor. We experimentally compare uniform nearest-neighbour (NN) and engineered perfect state transfer coupling profiles, benchmarking both against exact numerical simulations and demonstrating the superior state-transfer fidelity achieved through Hamiltonian engineering. Building on these results, we demonstrate how the commutation structure of the spin Hamiltonian can be exploited to realize a parallel digital implementation with reduced circuit depth and a more faithful approximation to the target dynamics. Our work establishes programmable trapped-ion processors as a powerful platform for exploring engineered spin dynamics and lays the groundwork for efficient quantum communication in future large-scale quantum technologies.

\section{Spin chain dynamics and digital implementation}

Quantum state transfer in spin chains provides a natural mechanism for communicating quantum information between spatially separated qubits through the intrinsic dynamics of interacting many body systems \cite{Bose2003}. Throughout this work we consider a one dimensional spin chain of $N$ qubits described by the XY Hamiltonian,

\begin{equation}
H =
\sum_{i=1}^{N}\sum_{j=i+1}^{N}
J_{ij}
\left(
X_iX_j + Y_iY_j
\right),
\label{eq:H_exact}
\end{equation}

where $X_i$ and $Y_i$ denote the Pauli operators acting on qubit $i$, and $J_{ij}$ is the exchange coupling between qubits $i$ and $j$. We consider ferromagnetic couplings ($J_{ij}<0$), for which the Hamiltonian conserves the total excitation number, making it a convenient model for quantum state transfer \cite{Gier2017}. The system evolves according to the unitary operator

\begin{equation}
U(t)=e^{-iHt},
\label{eq:time_evolution}
\end{equation}

such that

\begin{equation}
|\psi(t)\rangle = U(t)|\psi(0)\rangle,
\end{equation}

where $|\psi(0)\rangle$ and $|\psi(t)\rangle$ denote the initial and evolved states, respectively.

The transport properties are determined by the coupling profile $J_{ij}$. We first consider uniform nearest-neighbour (NN) couplings,

\begin{equation}
J_{ij}=
\begin{cases}
J, & |i-j|=1,\\
0, & \text{otherwise},
\end{cases}
\label{eq:NNcouplings}
\end{equation}

which represent the simplest realization of a spin chain.
Next we consider the engineered coupling profile proposed by Christandl \textit{et al.} \cite{Christandl2004},

\begin{equation}
J_j
=
J_{\mathrm{0}}
\sqrt{j(N-j)},
\qquad
j=1,\ldots,N-1,
\label{eq:PSTcouplings}
\end{equation}

which produces an equally spaced energy spectrum and enables perfect state transfer (PST) in ideal spin chains. In this work we compare the performance of uniform and engineered coupling profiles using both numerical simulations and experiments conducted on an IonQ trapped-ion quantum processor.

To experimentally realize the spin chain dynamics, we digitally simulate the Hamiltonian in Eq.~(\ref{eq:H_exact}) on IonQ's Forte 1/Forte Enterprise 1 trapped-ion processor, depending on the availability. Since the processor is gate-based, the continuous time evolution generated by Eq.~(\ref{eq:time_evolution}) is approximated using the first order Lie-Trotter product formula \cite{Trotter1959,Hall2015Ch2},

\begin{equation}
U(t)
=
e^{-iHt}
\approx
\left(
\prod_{k=1}^{m}
e^{-iH_k\Delta t}
\right)^n,
\label{eq:1st_ord_Trotter}
\end{equation}

where $H=\sum_{k=1}^{m}H_k$, $\Delta t=t/n$, and $n$ is the number of Trotter steps.

\begin{figure}[h]
    \centering
    \includegraphics[width=0.98\linewidth]{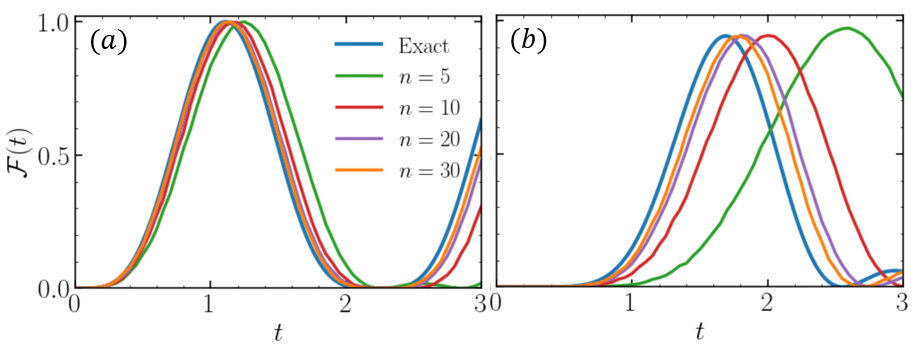}
    \caption{Convergence of the gate-based Trotter approximation towards the exact NN dynamics with increasing numbers of Trotter steps for (a) $N=3$, and (b) $N=5$. As the number of Trotter steps increases, the gate-based evolution approaches the exact Hamiltonian evolution.}
    \label{fig2}
\end{figure}

\begin{figure*}
    \centering
    \includegraphics[width=1.\linewidth]{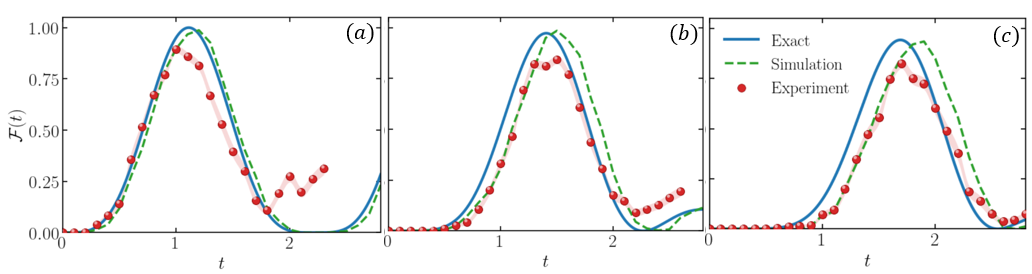}
    \caption{Experimental state-transfer fidelity (red circles) as a function of time for NN spin chains with uniform coupling strength $J=-1$, compared with the corresponding gate-based simulations (green dashed lines) and exact Hamiltonian evolution (blue solid lines) for (a) $N=3$, (b) $N=4$, and (c) $N=5$. As the chain length increases, the fidelity peak shifts to later times while its maximum value decreases, reflecting the increasing challenge of coherent state transfer over longer spin chains. Nevertheless, the measured fidelities remain above the classical threshold of $2/3$ for all system sizes investigated, demonstrating a quantum advantage. Both the experimental and gate-based results were obtained using 400 measurement shots, with the red shaded regions indicating the corresponding statistical uncertainties. The $N=3$ experiment was performed on IonQ's Forte Enterprise 1 processor, whereas the $N=4$ and $N=5$ results were obtained using IonQ's Forte 1 processor.}
    \label{fig3}
\end{figure*}

For the XY Hamiltonian, each interaction term acts on a pair of qubits,

\begin{equation}
H_{ij}
=
J_{ij}
\left(
X_iX_j+Y_iY_j
\right),
\end{equation}

giving the short time evolution

\begin{align}
e^{-iH_{ij}\Delta t}
&=
e^{-iJ_{ij}(X_iX_j+Y_iY_j)\Delta t}
\nonumber\\
&=
e^{-iJ_{ij}X_iX_j\Delta t}
e^{-iJ_{ij}Y_iY_j\Delta t},
\label{eq:pair_split}
\end{align}

where the two interaction terms commute. The full evolution is therefore approximated by

\begin{align}
U(t)
\approx
\left[
\prod_{i=1}^{N}
\prod_{j=i+1}^{N}
e^{-iJ_{ij}X_iX_j\Delta t}
e^{-iJ_{ij}Y_iY_j\Delta t}
\right]^n.
\label{eq:Ut_full_pairwise}
\end{align}

The trapped-ion architecture of IonQ is particularly well suited to this implementation because the required $XX$ and $YY$ interaction terms can be efficiently synthesized from the their native gates set with relatively low compilation overhead. The corresponding two-qubit rotations are

\begin{equation}
R_{XX}^{(i,j)}(\theta_{ij})
=
e^{-i\frac{\theta_{ij}}{2}(X_i\otimes X_j)},
\qquad
R_{YY}^{(i,j)}(\theta_{ij})
=
e^{-i\frac{\theta_{ij}}{2}(Y_i\otimes Y_j)},
\label{eq:XXYYUnitary}
\end{equation}

where $\theta_{ij}=2J_{ij}\Delta t$. Consequently, one Trotter step is implemented as

\begin{equation}
U(t)
\approx
\left[
\prod_{i=1}^{N}
\prod_{j=i+1}^{N}
R_{YY}^{(i,j)}(\theta_{ij})
R_{XX}^{(i,j)}(\theta_{ij})
\right]^n,
\label{eq:Trotter_native}
\end{equation}

whose repeated application reproduces the spin chain dynamics with increasing accuracy as the number of Trotter steps is increased, while enabling the engineered coupling profiles to be implemented directly on the trapped-ion processor.

In most of the numerical simulations and the trapped-ion experiments we perform, quantum state transfer is realised through the propagation of a single spin excitation along the chain. The system is initialised with the excitation localised on the first site,
\begin{equation}
|\psi_i\rangle = |100\cdots0\rangle,
\end{equation}
corresponding to the first qubit in the state $\ket{1}$ and all remaining qubits in the state $\ket{0}$. Under the evolution generated by the XY Hamiltonian, this excitation propagates coherently through the spin chain while conserving the total excitation number. The objective of the communication protocol is to transfer the excitation to the opposite end of the chain, such that at the optimal transfer time the system reaches the target state
\begin{equation}
|\psi_f\rangle = |000\cdots001\rangle.
\end{equation}

The performance of the communication protocol is quantified using the state transfer fidelity,
\begin{equation}
F(t)
=
\left|
\langle\psi_f|\psi(t)\rangle
\right|^2,
\label{eq:fidelity}
\end{equation}
where $|\psi(t)\rangle$ is the time-evolved state of the system and $|\psi_f\rangle$ denotes the target state corresponding to perfect transfer. Unless otherwise stated, the fidelity is evaluated at the optimal transfer time determined from numerical simulations. A fidelity of unity corresponds to perfect quantum state transfer, while the classical benchmark is $2/3$~\cite{Bose2007}. Fidelities exceeding this threshold therefore demonstrate genuine quantum communication beyond the classical limit.

\section{Experimental Demonstration of communication protocols}
\label{results}
To determine an appropriate number of Trotter steps for the experimental implementation, we first examine the convergence of the first-order Trotter approximation with increasing discretization of the time evolution. Figs.~\ref{fig2} compare the complete fidelity evolution obtained from the exact Hamiltonian with the Trotterized evolution for a three-qubit and five-qubit chain. As expected, the approximation systematically approaches the exact dynamics with increasing Trotter number, although the convergence becomes slower as the chain length increases.

Increasing the number of Trotter steps reduces the discretization error but simultaneously increases the circuit depth and, consequently, the number of entangling gates executed on the trapped-ion processor. The Trotter step number must therefore balance simulation accuracy against the accumulation of hardware errors arising from deeper circuits. Throughout this work we employ 30 Trotter steps, which provide converged transfer fidelities and transfer times for the system sizes considered whilst keeping the circuit depth sufficiently low for reliable experimental implementation.
The statistical uncertainty arising from finite sampling was estimated assuming binomial counting statistics. For a measured probability $p$, obtained from $N_{\mathrm{shots}}$ measurement shots, the corresponding uncertainty is given by
\[
\sigma=\sqrt{\frac{p(1-p)}{N_{\mathrm{shots}}}},
\]
where $p$ denotes the experimentally measured probability (or fidelity) associated with the corresponding measurement outcome.

\subsection{Experimental Results For Nearest-Neighbour Coupling}

The experimentally measured state-transfer fidelities for uniform nearest-neighbour spin chains with $N=3$, $4$, and $5$ qubits are shown in Fig.~\ref{fig3}. Overall, the experimental results closely reproduce the expected dynamics predicted by both the exact Hamiltonian evolution and the gate-based simulations, demonstrating that digitally simulated spin chains can be faithfully implemented on a trapped-ion quantum processor. The measured fidelity peaks occur at the expected evolution times and remain well above the classical threshold of $2/3$ for all system sizes investigated, thereby demonstrating successful quantum state transfer. The statistical uncertainty arising from finite sampling is indicated by the light red shaded regions.

As the chain length increases, the transfer fidelity gradually decreases while the fidelity peak shifts to later evolution times, reflecting the increasing time required for the excitation to propagate across longer spin chains. This trend is consistent with the theoretical behaviour of uniformly coupled nearest-neighbour spin chains and demonstrates that the trapped-ion processor accurately captures the underlying Hamiltonian dynamics. The small reduction in the experimentally measured fidelities relative to the ideal simulations is primarily attributed to finite gate fidelities, decoherence, and readout errors accumulated during the digital implementation.

Although the agreement between theory and experiment demonstrates that nearest-neighbour spin chains can be reliably realized on current trapped-ion hardware, the observed reduction in fidelity with increasing chain length indicates that uniform couplings are not optimal for scalable quantum communication. This naturally motivates the use of engineered coupling profiles that suppress dispersion and enhance state-transfer fidelity. In the following section, we investigate the engineered PST protocol introduced by Christandl \textit{et al.}~\cite{Christandl2004} and experimentally demonstrate its advantages over the uniform nearest-neighbour model.

\subsection{Experimental Results for Engineered Optimal Coupling}

\begin{figure}[h]
    \centering
    \includegraphics[width=0.93\linewidth]{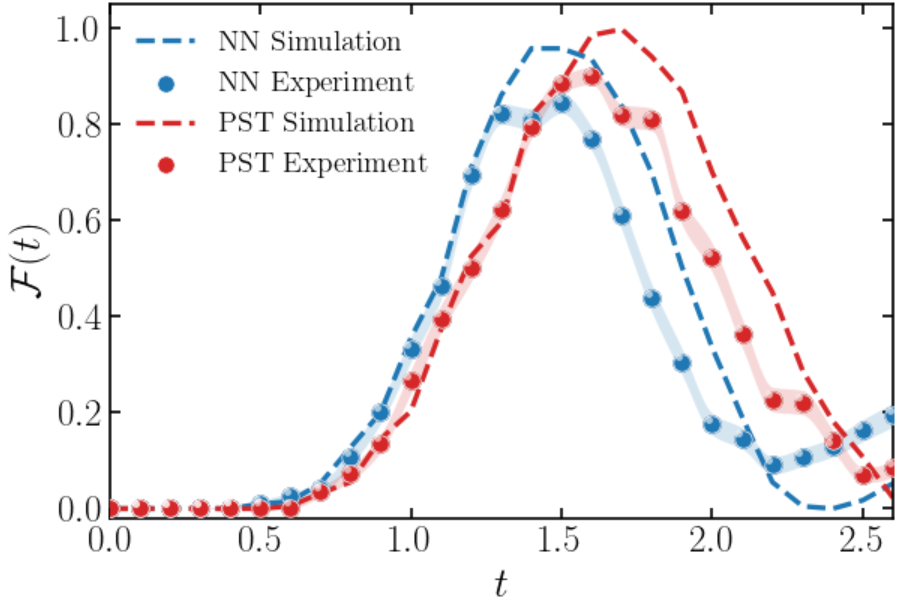}
   \caption{Comparison of the state-transfer fidelity for a four-qubit spin chain with uniform NN and engineered PST coupling profiles. The blue and red dashed curves correspond to the gate-based simulations for the NN and PST Hamiltonians, respectively, while the blue and red circles denote the corresponding experimental measurements obtained on IonQ's Forte-1 trapped-ion quantum processor. The NN chain employs uniform couplings with $J=-1$, whereas the PST chain uses the engineered coupling profile $J_n=J_0\sqrt{n(N-n)}$ with $J_0=-1$. The experimental results faithfully reproduce the enhancement in state-transfer fidelity achieved through Hamiltonian engineering, demonstrating the superiority of the engineered PST coupling over the uniform nearest-neighbour chain.}
    \label{fig5}
\end{figure}

The engineered coupling profile introduced by Christandl \textit{et al.} ~\cite{Christandl2004} provides a route to overcoming the degradation in fidelity observed for uniform nearest-neighbour spin chains. The coupling strengths,

\[
J_j = J_{\mathrm{0}}\sqrt{j(N-j)},
\]

produce an equally spaced energy spectrum that suppresses dispersion and enables coherent transport of the excitation across the chain with perfect fidelity in the ideal case. Consequently, unlike uniform nearest-neighbour couplings, the engineered Hamiltonian realizes perfect state transfer irrespective of the chain length, establishing it as an optimal protocol for quantum state transfer.

The numerical and experimental results are compared in Fig.~\ref{fig5} for $N=4$ which demonstrates experimental realization for a four-qubit chain on the trapped-ion quantum processor with engineered coupling together with the corresponding gate-based simulations. The measured dynamics closely reproduce the theoretical predictions, with the PST coupling exhibiting a higher transfer fidelity than the uniform NN chain. This difference become more pronounced with increasing system sizes. Although the experimentally observed fidelity is reduced by finite gate fidelities, decoherence, and other hardware imperfections, the enhancement provided by the engineered coupling is clearly preserved. These results demonstrate that digitally simulated spin chains on current trapped-ion hardware faithfully capture the advantages of engineered coupling and provide an experimentally viable route towards high-fidelity quantum communication.

\subsection{Parallel Trotterization of the XX+YY Spin Chain}

Having demonstrated that engineered coupling profiles enable high-fidelity quantum state transfer, we next investigate how the corresponding Hamiltonian can be implemented more efficiently on a digital quantum computer. In the conventional approach, each nearest-neighbour interaction is applied sequentially within every Trotter step. Although this faithfully reproduces the target Hamiltonian in the limit of many Trotter steps, it does not exploit the underlying structure of the Hamiltonian and therefore results in an unnecessarily deep quantum circuit.

\begin{figure}[h]
    \centering
    \includegraphics[width=0.98\linewidth]{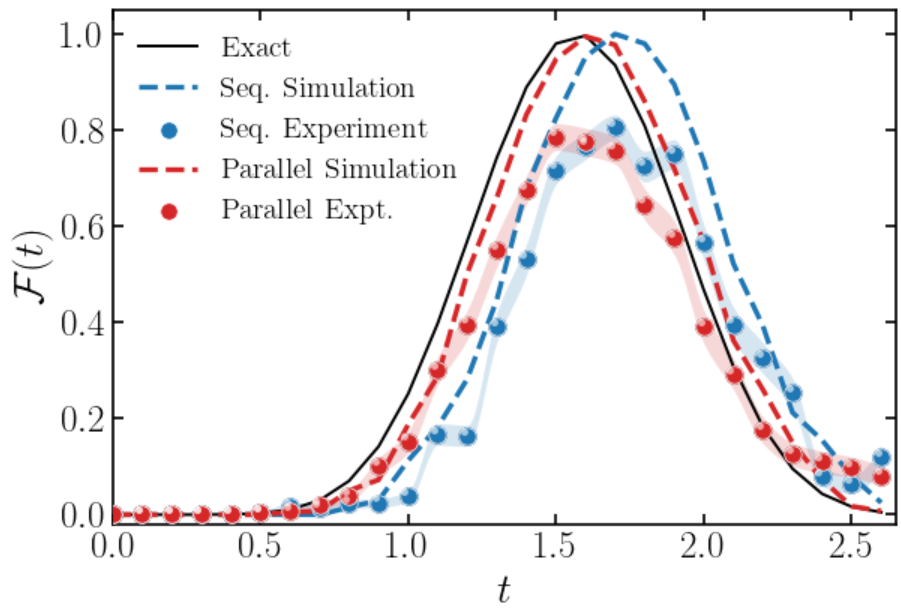}
   \caption{Comparison of the sequential and parallel implementations of the engineered perfect state transfer (PST) Hamiltonian for $N=5$. The solid black curve shows the exact continuous time evolution, while the blue and red dashed curves correspond to ideal gate-based simulations using the sequential and parallel Trotter decompositions, respectively. The circles denote the corresponding experimental results obtained on IonQ's Forte Enterprise processor. The parallel implementation reaches the maximum state transfer fidelity at an earlier evolution time while maintaining a comparable peak fidelity, thereby reducing the circuit depth and overall execution time by exploiting the parallel application of commuting interactions. Furthermore, both the simulated and experimental results show that the parallel decomposition more closely follows the exact dynamics than the sequential implementation.}
    \label{fig7}
\end{figure}

The nearest-neighbour XX+YY Hamiltonian possesses a simple commutation structure that can be exploited to improve the digital implementation. Interaction terms acting on disjoint pairs of qubits commute exactly and may therefore be applied simultaneously. For example, the interactions acting on the bonds $(1,2)$ and $(3,4)$ commute because they operate on different qubits, whereas neighbouring interactions such as $(1,2)$ and $(2,3)$ share a common qubit and consequently do not commute. The Hamiltonian can therefore be partitioned into two commuting sets corresponding to the odd and even bonds of the chain. A first-order Suzuki--Trotter decomposition then naturally implements each Trotter step using only two successive entangling layers, one containing all odd bonds and the other all even bonds.

This decomposition offers two advantages. By grouping commuting interaction terms within the same Trotter layer, it provides a more faithful digital realization of the target Hamiltonian for a fixed Trotter depth, reducing the error associated with the sequential ordering of non-commuting interactions. At the same time, because all disjoint interactions within a given layer may be executed simultaneously, the circuit depth is substantially reduced. Whereas a conventional implementation requires $N-1$ sequential two-qubit interactions per Trotter step for an $N$-qubit chain, the even-odd decomposition requires only two entangling layers irrespective of the system size. Consequently, the advantage of the parallel implementation becomes increasingly significant as the chain length increases.

These improvements are demonstrated in Fig.~\ref{fig7}, which compares the conventional sequential implementation with the parallel decomposition for an engineered perfect state transfer chain. The solid black curve shows the exact Hamiltonian evolution, while the dashed curves correspond to the ideal gate-based simulations and the circles denote the experimental measurements. The parallel implementation follows the exact dynamics more closely than the sequential approach and reaches the maximum state-transfer fidelity at an earlier evolution time while maintaining a comparable peak fidelity. The experimental results reproduce the same behaviour, confirming that exploiting the commutation structure of the Hamiltonian simultaneously improves the accuracy of the digital simulation and reduces the circuit depth required to implement the spin-chain dynamics.

A natural question is whether the quantum state could instead be transported through a sequence of nearest-neighbour SWAP operations. Such a routing protocol transfers the state exactly between adjacent qubits and therefore does not rely on a Trotter approximation. However, it constitutes a fundamentally different approach in which the quantum information is relayed serially through every intermediate site, requiring $N-1$ successive transport operations to traverse an $N$-qubit chain. Since a SWAP gate itself is typically synthesized from multiple native entangling operations (or equivalently requires a finite interaction time under an exchange Hamiltonian), the resulting circuit depth increases linearly with the transport distance.

One may further ask whether the engineered PST coupling profile could instead be incorporated into such a SWAP-based protocol by assigning bond-dependent interaction strengths
\[
J_i = J_0\sqrt{i(N-i)}.
\]
In this hypothetical scenario, the total transport time becomes
\[
T_{\rm SWAP}
=
\sum_{i=1}^{N-1}\frac{\pi}{4J_i}
=
\frac{\pi}{4J_0}
\sum_{i=1}^{N-1}
\frac{1}{\sqrt{i(N-i)}}
\underset{N\to\infty}{\simeq}
\frac{\pi^2}{4 J_0}.
\]
whereas Hamiltonian-mediated perfect state transfer occurs after
\[
T_{\rm PST}=\frac{\pi}{2J_0},
\]
up to the convention adopted for the XX+YY interaction strength. Thus, even under the same engineered coupling profile, the SWAP protocol remains slower by a constant prefactor of approximately $\pi/2$. More importantly, Hamiltonian-mediated transport exploits the collective dynamics of the entire spin chain, with all couplings acting simultaneously, whereas SWAP routing proceeds through a sequence of discrete nearest-neighbour exchanges.

\begin{figure}
    \centering
    \includegraphics[width=0.93\linewidth]{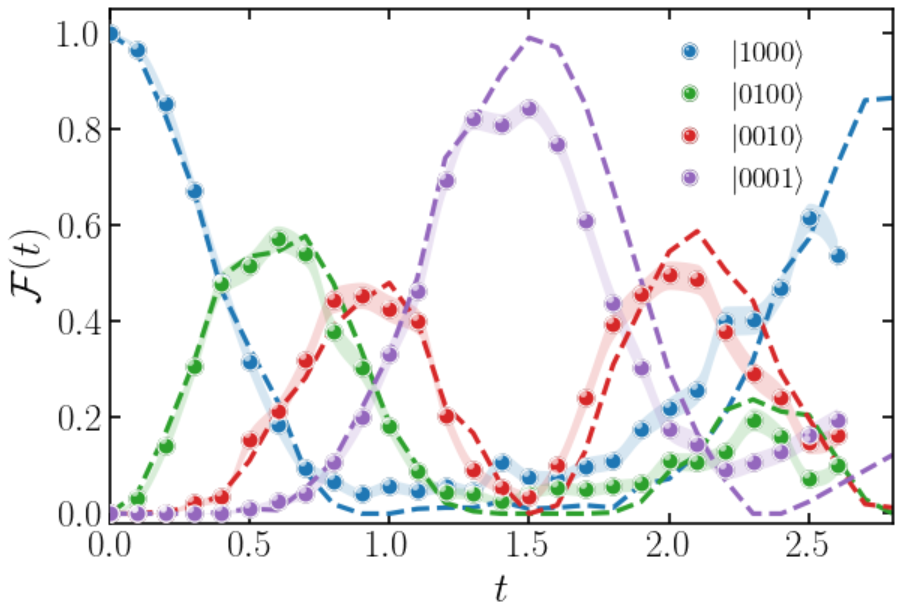}
    \caption{Time evolution of the populations of all single-excitation basis states for $N=4$ having nearest-neighbour coupling with $J=-1$. The dashed curves denote the gate-based simulations, while the circles represent the corresponding experimental measurements obtained on IonQ's Forte-1 trapped-ion quantum processor. The excitation is initially localized on the leftmost qubit (blue) and is transported to the rightmost qubit (purple) through the coherent population of the intermediate basis states (green and red), illustrating the intrinsic spin-chain dynamics responsible for quantum state transfer.}
    \label{fig4}
\end{figure}

To gain further insight into the mechanism underlying the quantum state transfer demonstrated in Section~\ref{results}, we examine the time evolution of the populations of all single-excitation basis states. Figure~\ref{fig4} shows these dynamics for a four-qubit nearest-neighbour spin chain. At $t=0$, the excitation is completely localized on the leftmost qubit, corresponding to the initial state (blue), which has unit fidelity and confirms the successful initialization of the quantum processor. As the system evolves under the spin Hamiltonian, the population of the initial state gradually decreases while the population of the target state (purple), corresponding to the excitation on the rightmost qubit, increases. The transfer is not instantaneous; rather, the excitation propagates coherently through the chain, giving rise to finite populations of the intermediate basis states (green and red) before arriving at the final qubit. This clearly demonstrates that quantum state transfer is mediated by the intrinsic many-body dynamics of the engineered spin chain, with the excitation propagating coherently through the intermediate qubits rather than being transferred directly between the two ends of the chain. 

\section{Leakage error and Hardware Limitations}
One of the principal limitations of the digital trapped-ion simulations is the accumulation of hardware errors during circuit execution. The dominant contribution arises from gate imperfections that change the total number of excitations in the system, thereby violating the excitation-number conservation of the ideal XY Hamiltonian. These errors cause the quantum state to leak from the intended single-excitation subspace into states containing either zero or multiple excitations.

\begin{figure}[h]
    \centering
    \includegraphics[width=0.98\linewidth]{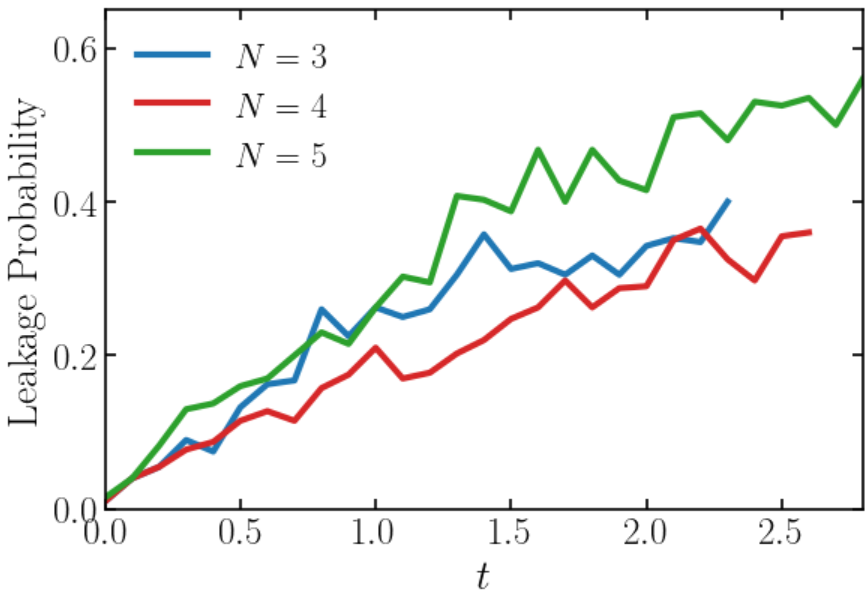}
    \caption{Leakage probability versus evolution time for nearest-neighbour coupled spin chains with $N=3,4,5$ qubits. $N=3$ was done on the IonQ's Forte Enterprise (in Jan 2026) and $N=4,5$ was done on IonQ's Forte-1 (in March 2026) quantum processor. Here, the leakage probability is defined as the population outside the single-excitation subspace. If we compare $N=4$ and $N=5$, the experiments for which were performed around the same time on the same machine, we find that the leakage increases with system size as expected. However, for $N=3$ we find that the error is more than $N=4$, thus showing that in Jan 26 the Forte Enterprise was worse that Forte-1. the For all of them the error increases with time demonstrating the accumulation of hardware errors as the circuit depth grows.}
    \label{fig8}
\end{figure}

For a three-qubit chain, the ideal evolution is restricted to the single-excitation basis
\begin{equation}
\{\ket{\psi_1},\ket{\psi_2},\ket{\psi_3}\}
=
\{\ket{100},\ket{010},\ket{001}\},
\end{equation}
whereas outcomes such as
\begin{equation}
\ket{000},\,
\ket{011},\,
\ket{101},\,
\ket{110},\,
\ket{111}
\end{equation}
represent leakage into unintended excitation sectors. Since the ideal dynamics conserve the total excitation number, these outcomes were discarded during the data analysis. The probabilities of all single-excitation states were first summed to obtain the total probability remaining within the logical subspace, after which the probabilities of the individual logical states were renormalised by this quantity. This post-selection removes leakage events and enables a direct comparison between the experimental results and the ideal single-excitation dynamics.

To quantify the extent of this effect, we define the leakage probability as
\begin{equation}
P_{\mathrm{leak}}
=
1-
\sum_{i\in\mathrm{SES}}P_i,
\end{equation}
where the summation is taken over all basis states belonging to the single-excitation subspace (SES). Figure~\ref{fig8} shows the measured leakage probability as a function of evolution time for spin chains containing three, four and five qubits. In all cases, the leakage probability increases with evolution time, reflecting the accumulation of gate errors as the circuit depth increases.

A comparison of the four- and five-qubit data, both obtained on the IonQ Forte 1 processor in March 2026, further demonstrates that the leakage probability increases with system size owing to the larger number of gates required to simulate longer spin chains. Interestingly, the experiment with $N=3$ exhibits a larger leakage probability than the $N=4$ qubit case, despite requiring a shallower circuit. This experiment was performed earlier on the IonQ Forte Enterprise 1 processor (Jan 2026), whereas the $N=4$ and $N=5$ qubit experiments were carried out on the Forte-1 processor (March 2026). The larger leakage observed for the three-qubit system therefore reflects differences in hardware calibration and noise characteristics between the two processor generations, rather than an intrinsic dependence on system size. Consequently, the four- and five-qubit data provide the more meaningful comparison for understanding the scaling of hardware errors with circuit depth.

These observations highlight an important practical limitation when benchmarking quantum hardware. Experimental performance depends not only on the algorithm and circuit depth, but also on the specific processor, its calibration and its availability at the time of execution. Although using a single hardware platform would be preferable for systematic benchmarking, this is not always possible because of hardware availability and access constraints. As quantum processors continue to improve, comparisons between experiments performed on different processors should therefore be interpreted with caution.

\section{Discussion and Outlook}

In this work, we have demonstrated the digital implementation of engineered spin chain dynamics on a trapped-ion quantum computer and shown their applicability for quantum communication between distant qubits. Numerical simulations and hardware experiments consistently demonstrate that engineered coupling profiles significantly enhance the fidelity of quantum state transfer compared with uniform nearest-neighbour couplings, particularly as the communication distance increases. The experimental results closely reproduce the predicted dynamics despite the presence of finite gate fidelities and decoherence, establishing digitally simulated spin chains as a practical platform for investigating quantum communication protocols on current programmable quantum hardware.

Beyond achieving high-fidelity quantum state transfer, this work demonstrates that exploiting the underlying structure of the spin Hamiltonian enables more efficient digital implementations. The parallel even-odd Trotter decomposition simultaneously improves the accuracy of the digital simulation and reduces the circuit depth by executing commuting interactions in parallel. Together, these results establish Hamiltonian engineering and optimized digital simulation as powerful tools for realizing scalable quantum communication on future programmable quantum processors.

The work also opens several natural directions for future investigation. An immediate extension is the exploration of more general coupling profiles, including long-range interactions, to further optimize quantum communication and remote gate operations. It would also be interesting to investigate larger spin chains and more complex network geometries, where communication overhead becomes increasingly important for scalable quantum architectures. Finally, the same framework could be extended beyond single-excitation state transfer to study the distribution of multipartite entanglement and the implementation of more general remote quantum operations, providing further insight into the role of engineered spin dynamics in future quantum communication and distributed quantum computing.

\section*{Acknowledgments}
This project was funded and supported by the UK National Quantum Computer Centre [NQCC200921], which is a UKRI Centre and part of the UK National Quantum Technologies Programme (NQTP). MS and SB also acknowledge EPSRC-SFI funded project EP/X039889/1 (GeQuantumBus).

\bibliography{main}

\appendix

\section{Transfer of a Superposition State}

\begin{figure}[h]
    \centering
    \includegraphics[width=0.93\linewidth]{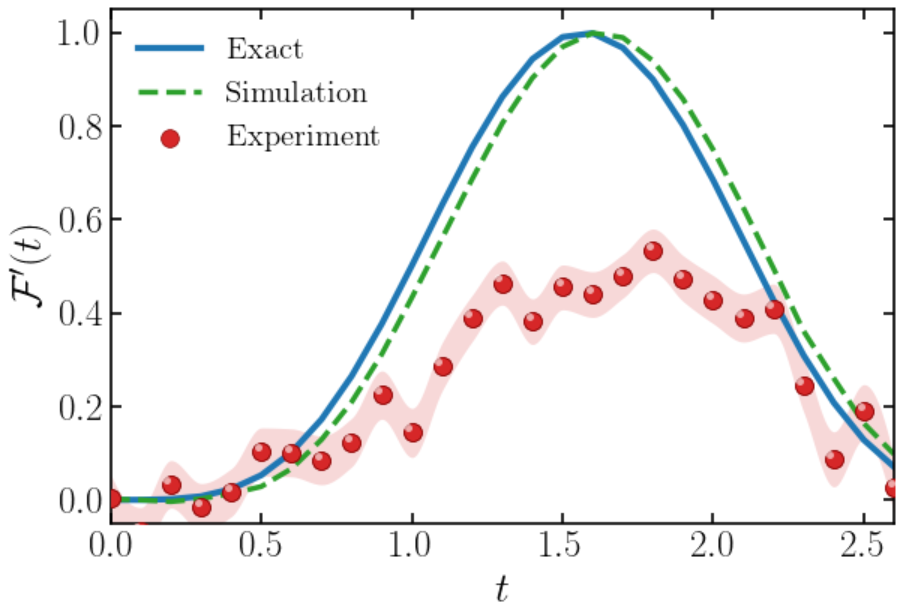}
    \caption{Time evolution of the superposition state transfer signal, $\mathcal{F}'(t)$, for the transfer of the state $|+\rangle=(|0\rangle+|1\rangle)/\sqrt{2}$. The solid blue and dashed green curves correspond to the exact Hamiltonian evolution and gate-based Trotter simulation, respectively, while the red circles denote the experimental measurements obtained on IonQ's Forte-1 trapped-ion quantum processor. }
    \label{figapp1}
\end{figure}

To demonstrate that the engineered spin chain preserves quantum coherence during transport, we also investigate the transfer of a superposition state. Starting from the ground state $|0\rangle^{\otimes N}$, a Hadamard gate is applied to the leftmost qubit, preparing the initial state,

\begin{equation}
|\psi(0)\rangle
=
|+\rangle\otimes|0\rangle^{\otimes(N-1)}
=
\frac{1}{\sqrt{2}}
\left(
|00\cdots00\rangle
+
|10\cdots00\rangle
\right),
\end{equation}

where

\begin{equation}
|+\rangle
=
\frac{|0\rangle+|1\rangle}{\sqrt{2}}.
\end{equation}

Ideally, the zero-excitation component remain unchanged under the XY Hamiltonian, while the single-excitation component is transported coherently across the spin chain by the engineered coupling profile. At the perfect state transfer time the ideal final state is,

\begin{equation}
|\psi_f\rangle
=
\frac{1}{\sqrt{2}}
\left(
|00\cdots00\rangle
+
|00\cdots01\rangle
\right)
=
|0\rangle^{\otimes(N-1)}
\otimes
|+\rangle,
\end{equation}

demonstrating that the quantum superposition has been transferred from the leftmost qubit to the end qubit.

To quantify the transfer, a Hadamard gate is applied to the $(N-1)$th (last) qubit immediately before measurement. For perfect state transfer, this decoding operation transforms the transferred superposition $|0\rangle^{\otimes(N-1)}
\otimes
|+\rangle$ into the $|00\cdots00\rangle$ state, enabling the transfer of the transported state to be inferred directly from computational-basis measurements without requiring full quantum state tomography. We therefore define the superposition state transfer signal as

\begin{equation}
\mathcal{F}'(t)
\equiv
\langle Z_{N-1}\rangle
=
P\!\left(q_{N-1}=0\right)
-
P\!\left(q_{N-1}=1\right),
\end{equation}

where $P(q_{N-1}=0)$ and $P(q_{N-1}=1)$ denote the probabilities of measuring the $(N-1)$th qubit in the states $|0\rangle$ and $|1\rangle$, respectively, after the final Hadamard operation. In the absence of errors, $\mathcal{F}'(t)$ evolves from zero at the initial time and reaches unity at the perfect state transfer time, providing a direct measure of the coherent transfer of the superposition state.

Figure~\ref{figapp1} compares the exact Hamiltonian evolution, the gate-based Trotter simulation, and the experimental measurements for the superposition state transfer signal, $\mathcal{F}'(t)$. The exact and gate-based results are in excellent agreement, confirming that the first-order Trotter decomposition accurately reproduces the coherent dynamics of the engineered spin chain. The experimental data exhibit the same qualitative time dependence, with the superposition transfer signal increasing towards the expected transfer time before decreasing again, demonstrating that the protocol preserves quantum coherence during the transport of an arbitrary single-qubit state.

The experimentally observed peak is noticeably lower than that predicted by the numerical simulations. While this reduction is consistent with the effects of finite gate fidelities, decoherence, and measurement errors, we believe that it is further influenced by the specific experimental conditions under which these data were acquired. In particular, the superposition state measurements were performed on IonQ's Forte-1 processor during the second week of July 2026, when the measured hardware performance was found to be significantly lower than that observed in our earlier experiments. This reduction is not unique to the superposition state protocol. During the same experimental time, the single excitation state-transfer experiments also exhibited substantially lower peak fidelities than those obtained using the same processor in March 2026. We therefore do not attribute the reduced peak in Fig.~\ref{figapp1} to the transfer of a superposition state itself, but rather to the overall experimental performance during that particular time.

At the time of writing, the Forte-1 processor is temporarily unavailable, preventing these measurements from being repeated immediately. We therefore expect to revisit this experiment once the processor becomes available again, with the aim of obtaining hardware results that more closely reproduce the theoretical predictions.

\end{document}